\documentclass[11pt]{article}

\usepackage[utf8]{inputenc}
\usepackage[T1]{fontenc}
\usepackage{amsmath,amssymb,amsfonts,amsthm}
\usepackage{graphicx}
\usepackage{geometry}
\usepackage[hidelinks]{hyperref}
\usepackage{authblk}

\title{\textbf{Long-Memory Estimation and Fractionally Integrated Modeling
		of White Maize Prices in Togo}}

\author[1]{Manganaw N'DAAM\thanks{Corresponding author: manganawn@gmail.com}}
\author[1]{Edoh KATCHEKPELE}
\author[1]{Tchilabalo Abozou KPANZOU}

\affil[1]{Laboratoire de Modélisation Mathématique et d'Analyse Statistique
	Décisionnelle (LaMMASD), Département de Mathématiques,
	Faculté des Sciences et Techniques, Université de Kara,
	BP 404 Kara, Togo}

\date{}

\begin{document}
	
	\maketitle
	
	\begin{abstract}
		Agricultural commodity prices often exhibit strong temporal persistence,
		which may limit the performance of conventional time series models.
		This study investigates long memory in logarithmic monthly white maize
		prices from six major markets in Togo between January 2001 and June 2022.
		Long memory is examined using the Geweke--Porter--Hudak, Local Whittle,
		Exact Local Whittle, and wavelet log-regression estimators. SARIMA,
		ARFIMA, and SARFIMA models are subsequently compared using the Bayesian
		Information Criterion and residual diagnostics. Long-range dependence
		is found across all markets. Fractionally integrated models provide the
		best fit for most markets, although SARIMA remains preferable for some.
		The results demonstrate that evidence of long memory does not necessarily
		imply that a fractionally integrated model provides the best empirical fit,
		emphasizing the importance of data-driven model selection.
	\end{abstract}
	
	\noindent\textbf{Keywords:}
	Long memory; Fractionally integrated processes; Exact Local Whittle;
	SARFIMA; White maize prices.
	
	\medskip
	
	\noindent\textbf{MSC 2020:} 62M10

\section{Introduction}

Agricultural commodity prices play a fundamental role in food security, producers' incomes, and the functioning of agricultural markets. Price fluctuations directly influence production, storage, and marketing decisions, while also affecting public policies aimed at ensuring market stability. Among the major staple crops in Sub-Saharan Africa, maize occupies a strategic position because of its importance in household consumption and its contribution to food security. Understanding the temporal dynamics of maize prices is therefore of considerable importance for producers, traders, and policymakers. Accordingly, recent studies continue to devote significant attention to agricultural price modeling, with particular emphasis on price volatility, persistence, and forecasting (\cite{devianto2024,meena2025,vorsah2025}).

Autoregressive integrated moving average (ARIMA) models and their seasonal extensions (SARIMA) constitute classical approaches to time series modeling \cite{box2015}. These models mainly describe short-range dependence, characterized by a relatively rapid decay of autocorrelations. However, many economic, financial, and environmental time series exhibit a much slower decay of their autocorrelation function, reflecting persistent dependence commonly referred to as long memory or long-range dependence (\cite{baillie1996,beran1994,granger1980,hosking1981}). Autoregressive fractionally integrated moving average (ARFIMA) models were specifically introduced to account for this persistence through a fractional differencing parameter. When long memory coexists with seasonal behavior, SARFIMA models allow both features to be incorporated simultaneously \cite{ray1993}.

Interest in fractionally integrated models remains strong in recent applied research. \cite{devianto2024} show, in the context of chili prices, that jointly accounting for long memory, seasonality, and exogenous variables can improve model performance. Similarly, \cite{meena2025} investigate the persistence of agricultural prices through a comparison of ARFIMA, ARIMAX, and ARFIMAX models. In a different application, \cite{qi2020} demonstrate the superiority of the SARFIMA model over the SARIMA model for a time series exhibiting both persistent dependence and seasonality. More recently, \cite{aruah2025} directly compared SARIMA and SARFIMA models for a seasonal long-memory process and concluded in favor of the SARFIMA specification within their application. Likewise, the study of \cite{vorsah2025} on maize price volatility in Sub-Saharan Africa highlights the continuing importance of statistical modeling for African cereal markets.

Despite these advances, several limitations remain in the existing literature. Studies on African agricultural markets still focus predominantly on price volatility, forecasting, or conventional ARIMA-type models, whereas systematic comparisons of SARIMA, ARFIMA, and SARFIMA models across multiple local markets remain relatively scarce. Furthermore, the characterization of long memory often relies on a single estimation method, although finite-sample properties may vary substantially across estimators. Semi-parametric methods, including the GPH estimator (\cite{geweke1983}), the Local Whittle estimator (\cite{robinson1995}), the Exact Local Whittle estimator (\cite{shimotsu2005}), and wavelet-based approaches (\cite{abry1998,moulines2007}), provide complementary tools for assessing the robustness of long-memory estimation. More importantly, evidence of long memory does not necessarily establish that a fractionally integrated model provides a better empirical representation than a conventional short-memory model. The identification of long memory and the selection of an appropriate time series model should therefore be regarded as related but distinct statistical questions. To the best of our knowledge, white maize prices observed in Togolese markets have not yet been investigated through a framework combining multiple semi-parametric estimators of long memory with a systematic comparison of classical and fractionally integrated models.

The objective of this study is therefore to investigate the temporal dynamics of monthly white maize prices observed in six major markets in Togo between January 2001 and June 2022. First, the long-memory parameter is estimated using four complementary approaches, namely the GPH estimator, the Local Whittle estimator, the Exact Local Whittle estimator, and wavelet log-regression. Second, three families of time series models, namely SARIMA, ARFIMA, and SARFIMA, are fitted separately to each market. The competing specifications are compared using the Bayesian
Information Criterion (BIC) proposed by \cite{schwarz1978}, and the
selected models are subsequently evaluated through residual
autocorrelation functions and the Ljung--Box test of \cite{ljung1978}. This framework addresses two related but distinct questions: whether long memory is present in white maize price series and whether evidence of long memory necessarily implies that a fractionally integrated model provides a better empirical representation than a conventional short-memory model.

The remainder of this paper is organized as follows.
Section~\ref{sec:data} describes the data and the corresponding preprocessing procedures.
Section~\ref{sec:methodology} presents the long-memory estimation methods together with the time series models considered in this study.
Section~\ref{sec:results} reports the empirical results, including the exploratory analysis, long-memory estimates, model comparison, and residual diagnostics.
Finally, Section~\ref{sec:conclusion} concludes the paper.

\section{Data}
\label{sec:data}

\subsection{Data source and study area}

This study uses monthly price series from the World Food Programme (WFP) food price database, made available through the Humanitarian Data Exchange (HDX) platform \cite{wfp_food_prices}. This database reports the prices of numerous food commodities observed in nearly 3,000 markets across 98 countries, with predominantly monthly series extending, depending on the country, back to the early 1990s.

Among the various commodities available, this study focuses exclusively on white maize, one of the main cereals consumed in Togo and a strategic commodity for household food security. The selected data correspond to monthly prices observed in the major Togolese markets between January 2001 and June 2022. Prices are expressed in CFA francs (XOF) per kilogram and constitute the variable of interest analyzed in this study.

Since the primary objective of this study is the characterization of long memory, particular attention was paid to the quality and length of the time series used. Indeed, semi-parametric long-memory estimation methods rely on asymptotic properties that require sufficiently long and regularly observed time series in order to produce stable and comparable estimates.

On this basis, only markets with an almost continuous monthly series over the entire common observation period were retained. The analysis therefore focuses on the markets of Korbongou, Cinkassé, Kara, Anié, Amégran, and Lomé. For these six markets, the observations cover the period from January 2001 to June 2022, corresponding to the last common period available for all selected series. Each market is therefore represented by a series of 258 monthly observations.

Beyond their temporal availability, these markets are of particular interest because they are distributed across the entire Togolese territory, from the northern regions to the capital city located in the south of the country. This geographical distribution makes it possible to investigate price dynamics under relatively diverse economic and agricultural conditions while maintaining a homogeneous analytical framework.

\subsection{Data preprocessing}

Before any statistical analysis, the dataset underwent a preprocessing stage designed to ensure the temporal consistency of the series and the reliability of the subsequent analyses. This step is particularly important in the context of long-memory analysis, as semi-parametric estimators are sensitive to irregular sampling frequencies, missing values, and discontinuities that may alter the dependence structure of the series.

A complete monthly calendar covering the period from January 2001 to June 2022 was first constructed for each of the six selected markets in order to verify the chronological continuity of the observations. This procedure revealed a very limited number of missing values, representing only 10 observations out of the 1,548 observations in the complete dataset (less than 1\% of the data), distributed over three distinct months: June 2019, November 2020, and December 2020. Given their small proportion and isolated occurrence, no observations were removed. The missing values were replaced using temporal interpolation, which exploits the chronological structure of the data in order to preserve the continuity of the series.

The impact of this procedure was subsequently assessed by comparing the main descriptive statistics before and after imputation. As shown in Table~\ref{tab:imputation_validation}, the changes observed in the measures of central tendency and dispersion remain very small for all the markets considered. These results indicate that the interpolation did not significantly modify the statistical properties of the series and confirm that the preprocessing stage does not introduce any bias likely to affect the subsequent analyses. The resulting dataset is therefore completely free of missing values.

The prices were subsequently transformed using the natural logarithm. This transformation is commonly employed in the analysis of economic and financial time series because it helps reduce distributional asymmetry, mitigate heteroscedasticity when present, and express price variations on a relative scale, thereby facilitating the interpretation of changes in terms of proportional variations. Consequently, all the analyses presented in the remainder of this paper are conducted using the logarithmic series. Table~\ref{tab:descriptive_statistics} reports the main descriptive statistics of the six monthly series after preprocessing. These results provide an initial overview of price levels and their variability before the stationarity analysis, long-memory estimation, and time series modeling presented in the following sections.

At the end of this preparation stage, the final dataset consists of six harmonized monthly series, each containing 258 observations, for a total of 1,548 regularly spaced observations over time. This dataset constitutes the basis for the exploratory analyses, long-memory estimation, and comparison of time series models presented in the following sections.

\begin{table}[htbp]
	\footnotesize
	\caption{\label{tab:imputation_validation}
		Effect of Temporal Interpolation on the Descriptive Statistics of the Selected Markets.}
	\centering
	
	\begin{tabular}{lccc}
		\hline
		\textbf{Market} &
		\textbf{Missing Values} &
		\textbf{Mean Change (\%)} &
		\textbf{Std. Dev. Change (\%)} \\
		\hline
		
		Korbongou & 1 & -0.0549 & -0.1542 \\
		Cinkassé  & 3 & -0.0835 & -0.5111 \\
		Kara      & 1 &  0.0226 & -0.1883 \\
		Anié      & 1 &  0.0053 & -0.1945 \\
		Amegnran  & 1 & -0.0286 & -0.1859 \\
		Lomé      & 3 &  0.0004 & -0.5843 \\
		
		\hline
		\textbf{Total} & \textbf{10} & -- & -- \\
		\hline
	\end{tabular}
	
	\vspace{0.15cm}
	
	\begin{minipage}{0.92\textwidth}
		\footnotesize
		\textit{Note:} Relative changes correspond to the percentage differences
		between the descriptive statistics computed before and after temporal interpolation.
	\end{minipage}
	
\end{table}

\begin{table}[htbp]
	\footnotesize
	\caption{\label{tab:descriptive_statistics}
		Descriptive Statistics of Monthly White Maize Prices in the Six Selected Markets (January 2001--June 2022).}
	\centering
	
	\begin{tabular}{lccccc}
		\hline
		\textbf{Market} &
		\textbf{Mean} &
		\textbf{Median} &
		\textbf{Std. Dev.} &
		\textbf{Min} &
		\textbf{Max} \\
		\hline
		
		Korbongou & 141.55 & 135.50 & 43.86 & 56.00 & 270.00 \\
		Cinkassé  & 142.60 & 135.00 & 42.71 & 57.00 & 274.00 \\
		Kara      & 147.45 & 152.50 & 47.27 & 50.00 & 282.00 \\
		Anié      & 133.19 & 130.00 & 48.02 & 43.00 & 269.00 \\
		Amégran   & 161.71 & 152.00 & 55.89 & 74.00 & 518.00 \\
		Lomé      & 189.27 & 186.00 & 56.33 & 75.00 & 350.00 \\
		
		\hline
	\end{tabular}
	
	\vspace{0.15cm}
	
	\begin{minipage}{0.92\textwidth}
		\footnotesize
		\textit{Note:} Statistics are computed from the final monthly dataset after
		temporal interpolation. Prices are expressed in CFA francs (XOF) per kilogram.
	\end{minipage}
	
\end{table}

\section{Methodology}
\label{sec:methodology}

\subsection{Theoretical foundations of long memory}

One of the fundamental characteristics of time series is the nature of their temporal dependence. In classical short-memory models, the influence of an observation on future observations decreases rapidly over time. In contrast, some series exhibit much more persistent dependence, characterized by a slow decay of temporal dependence. This phenomenon is generally referred to as long memory or long-range dependence.

From a theoretical perspective, a stationary process $\{X_t\}_{t\in\mathbb{Z}}$ is said to exhibit long memory when its autocovariance function
\[
\gamma(k)=\mathrm{Cov}(X_t,X_{t+k})
\]
decays according to a power law as the lag $k$ tends to infinity, that is,
\[
\gamma(k)\sim Ck^{2d-1}, \qquad k\rightarrow\infty,
\]
where $C>0$ is a constant and $d$ denotes the memory parameter. Unlike short-memory processes, whose autocovariances decay exponentially, this slow decay reflects the existence of persistent dependence between observations that are widely separated in time.

An equivalent characterization can be obtained in the frequency domain. When the process exhibits long memory, its spectral density has a singularity in the neighborhood of the zero frequency and satisfies
\[
f(\lambda)\sim C|\lambda|^{-2d},
\qquad
\lambda\rightarrow0,
\]
where $C$ is a strictly positive constant. Thus, long memory is manifested by a strong concentration of spectral power at very low frequencies, reflecting the presence of persistent long-term fluctuations.

The fractional parameter $d$ directly measures the intensity of this persistence. When $d=0$, the process does not exhibit long memory and behaves as a short-memory process. For $0<d<0.5$, the process remains stationary while exhibiting long-range dependence. When $0.5 \leq d < 1$, the process still exhibits long memory but becomes non-stationary. Negative values of $d$, on the other hand, indicate negative long-range dependence, also referred to as antipersistence.

In this study, the presence of long memory is assessed using four complementary semi-parametric estimators: the Geweke--Porter--Hudak (GPH) estimator, the Local Whittle estimator, the Exact Local Whittle estimator, and the wavelet log-regression estimator. These methods all rely on the asymptotic behavior of the spectrum at low frequencies but differ in their estimation principles and finite-sample properties. Their comparison makes it possible to assess the robustness of the estimated values before selecting the most appropriate estimator for the fractional modeling of the logarithms of white maize prices.

\subsubsection{Geweke and Porter-Hudak estimator}

The GPH estimator, proposed by \cite{geweke1983}, is one of the first semi-parametric methods developed to estimate the long-memory parameter of a fractionally integrated process. Unlike parametric approaches, it does not require the complete specification of the data-generating model and relies solely on the asymptotic behavior of the spectral density in the neighborhood of the zero frequency.

Under the assumption of a long-memory process, the spectral density satisfies
\[
f(\lambda)\propto |\lambda|^{-2d},
\qquad
\lambda\rightarrow0.
\]

Using the empirical periodogram
\[
I(\lambda_j)
=
\frac{1}{2\pi n}
\left|
\sum_{t=1}^{n}
X_t
e^{-i\lambda_j t}
\right|^2,
\]
computed at the Fourier frequencies
\[
\lambda_j=\frac{2\pi j}{n},
\qquad
j=1,\ldots,m,
\]
where $n$ denotes the sample size and $m$ the number of low frequencies retained for estimation, with $m \rightarrow \infty$ and $m/n \rightarrow 0$ as $n \rightarrow \infty$. The GPH estimator is then based on the linear regression
\[
\log\!\left(I(\lambda_j)\right)
=
\alpha
-
d
\log\!\left(
4\sin^2\frac{\lambda_j}{2}
\right)
+
\varepsilon_j,
\qquad
j=1,\ldots,m,
\]
where $\alpha$ denotes the intercept of the regression and $\varepsilon_j$ is a zero-mean error term. The memory parameter $d$ is obtained by ordinary least squares from the estimated slope of this regression. The choice of the number of frequencies $m$ plays a crucial role, as it determines the trade-off between bias and variance of the estimator. In practice, only the lowest frequencies are retained in order to satisfy the asymptotic approximation on which the method is based.

The GPH estimator has the advantage of being simple to implement and is widely used in the literature as a benchmark method for detecting long memory. However, its performance may be sensitive to the choice of the frequency band used as well as to the presence of short-memory components that are not accounted for in the regression.

\subsubsection{Local Whittle estimator}

The Local Whittle (LW) estimator, proposed by \cite{robinson1995}, is a semi-parametric alternative to the GPH estimator for estimating the long-memory parameter. Unlike the latter, it is not based on a linear regression of the logarithm of the periodogram but on the minimization of an approximate likelihood function constructed from the behavior of the spectral density in the neighborhood of the zero frequency. This approach generally provides greater statistical efficiency and increased robustness with respect to periodogram fluctuations.

Let $\lambda_j=2\pi j/n$, $j=1,\ldots,m$, denote the $m$ lowest Fourier frequencies, and let $I(\lambda_j)$ be the empirical periodogram of the series under study. The Local Whittle estimator is obtained by minimizing the objective function

\[
Q(d)
=
\log\left(
\frac{1}{m}
\sum_{j=1}^{m}
\lambda_j^{2d}
I(\lambda_j)
\right)
-
\frac{2d}{m}
\sum_{j=1}^{m}
\log(\lambda_j),
\]

where $d$ denotes the memory parameter to be estimated. The estimator is then defined as

\[
\widehat{d}_{LW}
=
\arg\min_{d}
Q(d).
\]

As with the GPH estimator, the choice of the number of frequencies $m$ directly influences the properties of the estimator. Using too few frequencies increases the variance of the estimator, whereas using too many frequencies may introduce bias by incorporating frequencies for which the asymptotic approximation is no longer valid.

The Local Whittle estimator offers several advantages. It is semi-parametric, does not require the complete specification of the underlying model, and possesses good asymptotic properties under relatively general conditions. Moreover, it is generally less sensitive than the GPH estimator to local fluctuations of the periodogram, which explains its widespread use in empirical studies of long-memory processes.

\subsubsection{Exact Local Whittle estimator}

The Exact Local Whittle (ELW) estimator, proposed by \cite{shimotsu2005}, is an extension of the Local Whittle estimator designed to improve the estimation of the memory parameter when the process may be either stationary or non-stationary. Whereas the classical Local Whittle estimator is primarily suited to stationary processes, the ELW approach relies on an exact formulation of the local likelihood that remains valid over a broader range of the memory parameter. This property yields an estimator that is consistent and asymptotically normal for a broad class of fractionally integrated processes.

The main idea is to apply a frequency-domain transformation that removes the effect of fractional differencing before estimating the memory parameter. Based on the modified discrete Fourier transform, the objective function is defined as

\[
R(d)
=
\log
\left(
\frac{1}{m}
\sum_{j=1}^{m}
I_{\Delta^{d}}(\lambda_j)
\right)
-
\frac{2d}{m}
\sum_{j=1}^{m}
\log(\lambda_j),
\]

where $I_{\Delta^{d}}(\lambda_j)$ denotes the periodogram of the series after correction for fractional differencing and $m$ represents the number of low frequencies retained for estimation.

The Exact Local Whittle estimator is then defined by

\[
\widehat{d}_{ELW}
=
\arg\min_{d}
R(d).
\]

Given its favorable asymptotic properties, particularly its consistency for both stationary and non-stationary processes, together with its good performance documented in the literature, the Exact Local Whittle estimator is now regarded as one of the benchmark methods for semi-parametric estimation of the long-memory parameter \cite{shimotsu2005}. For these reasons, it is retained in the remainder of this study to estimate the memory parameter used in the modeling of fractionally integrated processes.

\subsubsection{Wavelet log-regression estimator}

Wavelet-based methods provide an alternative framework for analyzing
long-range dependence by examining the behavior of wavelet coefficients
across scales \cite{percival2000}. In particular, wavelet log-regression
estimation exploits the asymptotic behavior of the variance of wavelet
coefficients at coarse scales. This approach was notably developed by
\cite{abry1998} and subsequently investigated within a general
semi-parametric framework by \cite{moulines2007}.

Let $\{W_{j,k}\}$ denote the wavelet coefficients obtained at scale $j$ and location $k$. For a process with memory parameter $d$, their variance satisfies, under appropriate conditions on both the process and the analyzing wavelet,

\[
\operatorname{Var}(W_{j,k})
\sim
C\,2^{2dj},
\qquad j\rightarrow\infty,
\]

where $C>0$ is a constant. Taking the base-two logarithm yields

\[
\log_2\!\left(\operatorname{Var}(W_{j,k})\right)
=
\alpha+2dj+o(1),
\]

where $\alpha=\log_2(C)$. The theoretical variance is replaced by the empirical scalogram

\[
\widehat{\sigma}_j^2
=
\frac{1}{n_j}
\sum_{k=1}^{n_j}W_{j,k}^2,
\]

where $n_j$ denotes the number of coefficients available at scale $j$. The memory parameter is then estimated from the slope of the linear regression

\[
\log_2\!\left(\widehat{\sigma}_j^2\right)
=
\alpha+\beta j+\varepsilon_j,
\]

according to

\[
\widehat d_{\mathrm{WLR}}
=
\frac{\widehat\beta}{2}.
\]

In this study, the long-memory parameter is estimated by ordinary least squares linear regression applied to the logarithms of the empirical variances of the wavelet coefficients over the selected scales.
\subsection{Time series models}

The main objective of this study is to determine whether accounting for long memory improves the modeling of monthly white maize prices compared with conventional approaches based on short-range dependence. To address this question, three families of models are considered. SARIMA models serve as the benchmark by assuming short memory and a possible seasonal component. ARFIMA models generalize ARIMA models by replacing integer differencing with fractional differencing in order to account for long memory. Finally, SARFIMA models extend this approach by simultaneously combining long memory and seasonality. The three models are briefly presented below.

\subsubsection{SARIMA model}

Seasonal Autoregressive Integrated Moving Average (SARIMA) models, developed by \cite{box2015}, are an extension of ARIMA models that simultaneously represent short-range dependence and seasonal fluctuations observed in a time series. They remain one of the most widely used approaches for modeling and forecasting economic and agricultural time series.

A SARIMA$(p,d,q)\times(P,D,Q)_s$ model is defined by

\[
\Phi(B^s)\phi(B)(1-B)^d(1-B^s)^DX_t
=
\Theta(B^s)\theta(B)\varepsilon_t,
\]

where $B$ denotes the backshift operator; $p$ and $q$ correspond to the non-seasonal autoregressive and moving average orders, respectively; $P$ and $Q$ denote the seasonal autoregressive and moving average orders; $d$ and $D$ are the non-seasonal and seasonal differencing orders; $s$ represents the seasonal period; and $\varepsilon_t$ is a white noise process with zero mean and constant variance. The polynomials $\phi(B)$, $\theta(B)$, $\Phi(B^s)$, and $\Theta(B^s)$ denote the non-seasonal and seasonal autoregressive and moving average polynomials, respectively.

In this study, the SARIMA model serves as the benchmark model based on the assumption of short memory. Its performance is compared with that of fractionally integrated models in order to assess the benefit of accounting for long-range dependence in the dynamics of white maize prices.

\subsubsection{ARFIMA model}

Autoregressive Fractionally Integrated Moving Average (ARFIMA) models, introduced independently by \cite{granger1980} and \cite{hosking1981}, generalize ARIMA models by allowing a fractional differencing order. This extension makes it possible to represent processes exhibiting long memory while preserving the autoregressive and moving average structure of conventional models.

An ARFIMA$(p,d,q)$ model is expressed as

\[
\phi(B)(1-B)^dX_t
=
\theta(B)\varepsilon_t,
\]

where the fractional parameter $d$ is no longer necessarily an integer. When $d=0$, the model reduces to a classical ARMA process, whereas positive values of $d$ indicate long-range persistence.

In this study, the memory parameter is estimated using the Exact Local Whittle estimator whenever the ARFIMA model is selected. This model therefore makes it possible to assess whether the introduction of fractional differencing improves the statistical representation of the series compared with the SARIMA model.

\subsubsection{SARFIMA model}

Seasonal Autoregressive Fractionally Integrated Moving Average (SARFIMA) models extend SARIMA models by replacing integer differencing with fractional differencing while preserving a seasonal structure. They therefore allow the simultaneous representation of short-range dependence, seasonal fluctuations, and long-memory effects.

The general formulation of a SARFIMA$(p,d,q)\times(P,D,Q)_s$ model is given by

\[
\Phi(B^s)\phi(B)
(1-B)^d
(1-B^s)^D
X_t
=
\Theta(B^s)\theta(B)\varepsilon_t,
\]

where the parameter $d$ now represents a fractional differencing order, while the remaining parameters retain the same interpretation as in the SARIMA model.

The SARFIMA model constitutes the most general framework considered in this study. It makes it possible to determine whether simultaneously accounting for long memory and seasonality significantly improves the fit of the white maize price series compared with the SARIMA and ARFIMA models.

\section{Results}
\label{sec:results}

\subsection{Exploratory analysis}

\begin{figure}[htbp]
	\centering
	\makebox{\includegraphics[width=0.95\textwidth]{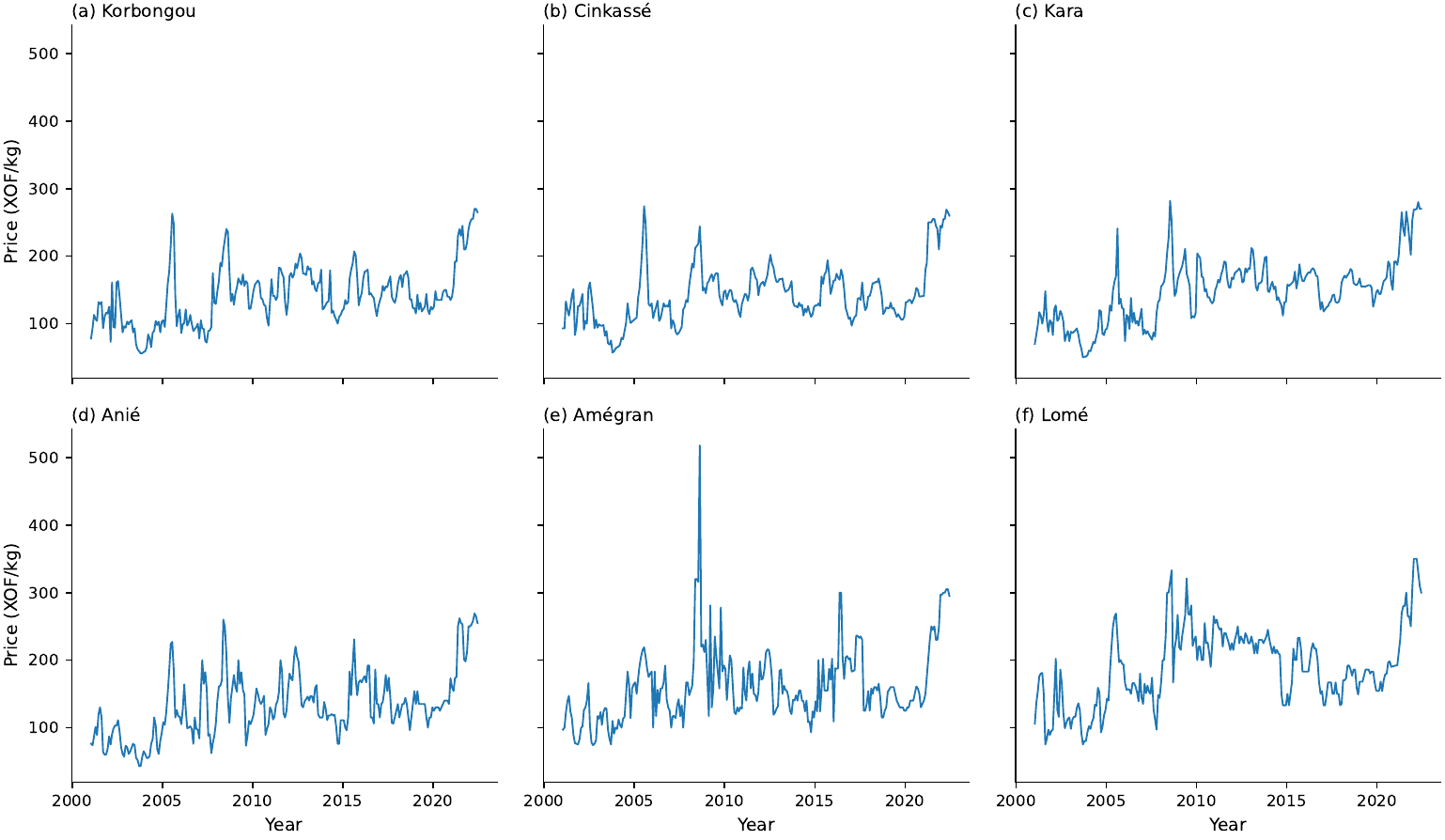}}
	\caption{\label{fig:price_series}Monthly White Maize Prices in the Six Selected Togolese Markets (January 2001--June 2022).}
\end{figure}

Figure~\ref{fig:price_series} presents the monthly evolution of white maize prices in the six selected markets between January 2001 and June 2022. Overall, the series exhibit an upward trend over the study period, reflecting a gradual increase in price levels over time. This overall evolution is, however, punctuated by fluctuations of varying magnitude across markets, reflecting the specific dynamics of each market.

Despite this common overall pattern, notable differences are observed across markets, both in terms of average price levels and variability. The markets located in the southern part of the country, particularly Lomé and Amégran, generally exhibit higher price levels than those observed in the northern regions. Episodes of sharp price increases are also visible in some series, suggesting the occurrence of occasional shocks likely to affect the balance between supply and demand.

The main descriptive statistics are reported in Table~\ref{tab:descriptive_statistics}. Mean prices range from 133.19 XOF/kg in Anié to 189.27 XOF/kg in Lomé, while the standard deviations range from 42.71 to 56.33 XOF/kg, highlighting heterogeneity in price volatility across markets. The Amégran market also exhibits the highest maximum value in the sample (518 XOF/kg), indicating the occurrence of exceptional price increases during the study period.

Overall, this descriptive analysis highlights marked spatial differences in the level and variability of white maize prices while revealing a common persistent temporal dynamics across the different markets. These preliminary findings justify a more in-depth investigation of the statistical properties of the series, particularly their stationarity and potential long-memory behavior.
\begin{figure}[htbp]
	\centering
	\makebox{\includegraphics[width=0.90\textwidth]{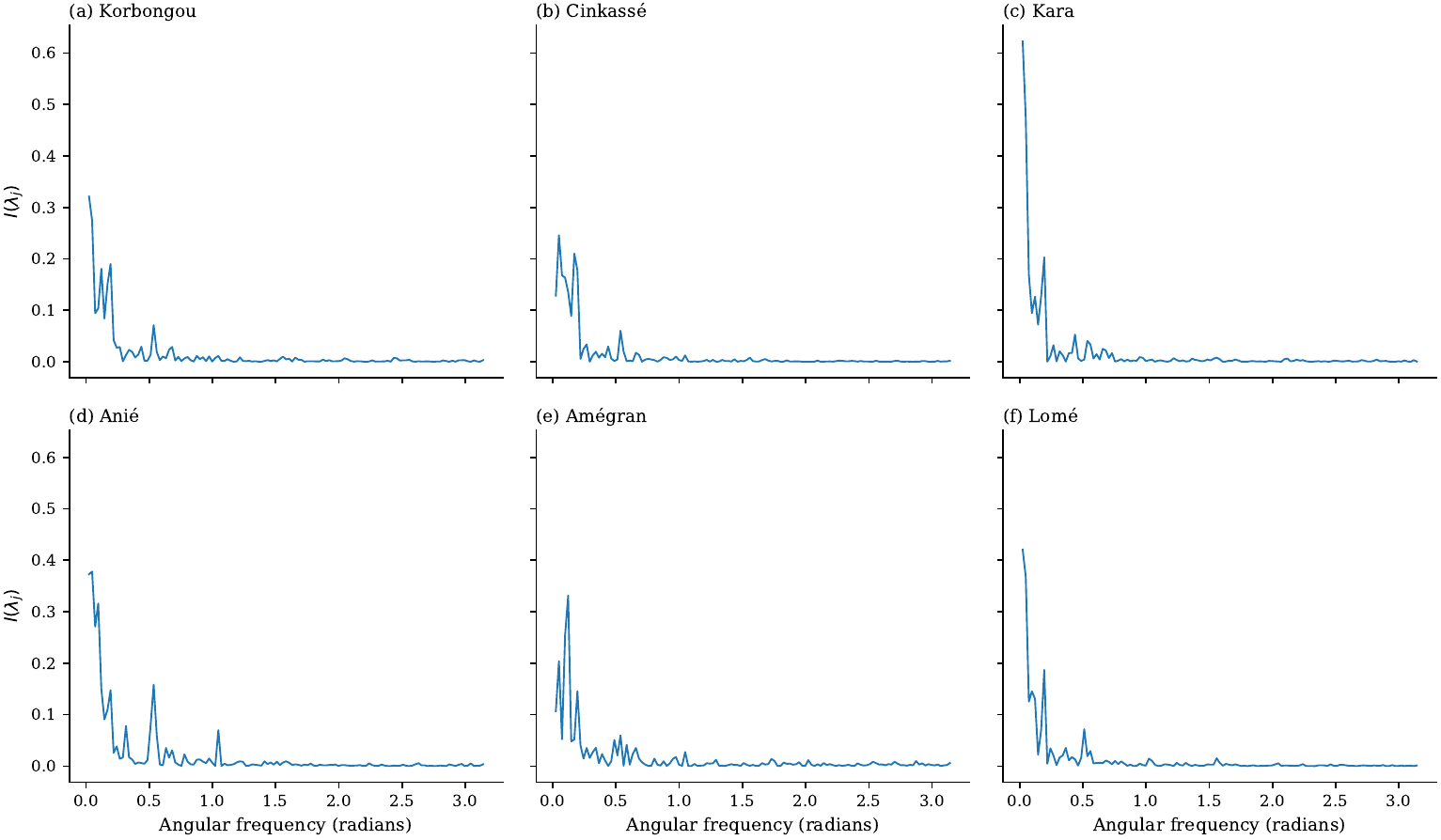}}
	\caption{\label{fig:periodograms}Empirical Periodograms of the Logarithmic Monthly White Maize Price Series for the Six Selected Markets.}
\end{figure}
Figure~\ref{fig:periodograms} presents the empirical periodograms of the logarithmic monthly white maize price series for the six selected markets. Across all markets, the spectral density appears to be highly concentrated in the neighborhood of the zero frequency and decreases rapidly as the frequency increases. This concentration of spectral power at low frequencies provides a first qualitative indication of temporal persistence that may reflect long-memory behavior. However, this observation remains descriptive and is not sufficient, by itself, to establish the existence of long-range dependence. The latter is therefore investigated quantitatively in the following sections using several semi-parametric estimators of the memory parameter.

\begin{figure}[htbp]
	\centering
	\makebox{\includegraphics[width=0.92\textwidth]{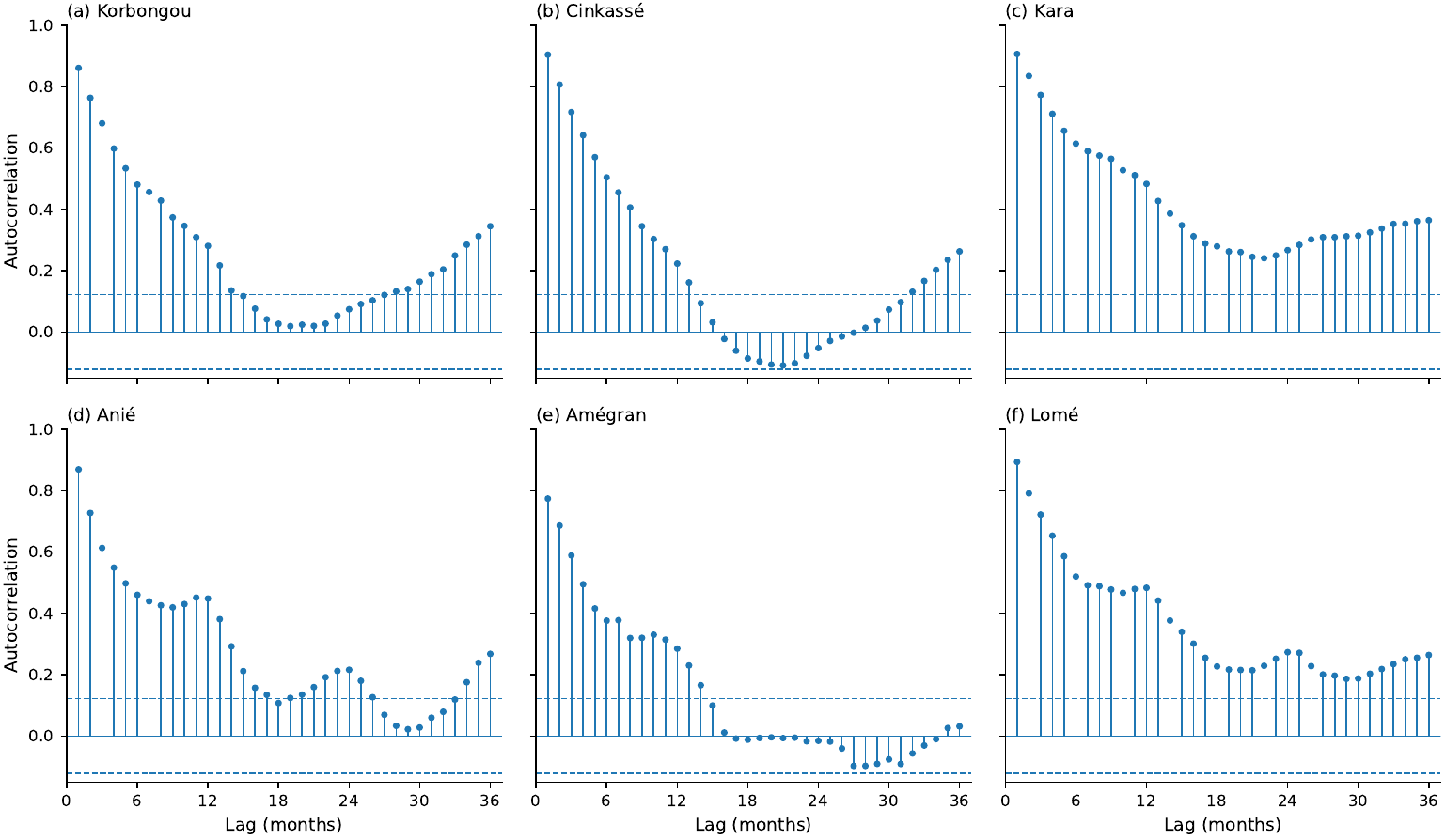}}
	\caption{\label{fig:acf}Sample Autocorrelation Functions of the Logarithmic Monthly White Maize Prices for the Six Selected Markets.}
\end{figure}

Figure~\ref{fig:acf} presents the empirical autocorrelation functions of the logarithmic monthly white maize price series for the six markets under study. Across all markets, the autocorrelations remain positive over a large number of lags and decay gradually rather than disappearing rapidly. This slow decay is a descriptive characteristic frequently observed in series exhibiting long-range dependence. The pattern is particularly pronounced for the Korbongou, Kara, and Lomé markets, where the autocorrelation coefficients remain well above the confidence bands for many lags. Although this observation does not constitute formal evidence of long memory, it suggests strong temporal persistence in the series and justifies the use of estimation methods specifically designed to quantify the memory parameter.
\subsection{Stationarity analysis}

The stationarity properties of the logarithmic price series were examined using the Augmented Dickey--Fuller (ADF) and Kwiatkowski--Phillips--Schmidt--Shin (KPSS) tests. The results are reported in Table~\ref{tab:stationarity_tests}.

The two tests do not lead to systematically consistent conclusions. For some markets, the ADF test rejects the null hypothesis of a unit root, whereas the KPSS test simultaneously rejects the null hypothesis of stationarity. For others, both tests suggest non-stationary behavior. This lack of consensus is frequently encountered in the analysis of economic time series exhibiting strong temporal persistence, as conventional stationarity tests have difficulty distinguishing a long-memory process from a process containing a unit root.

These results should therefore be interpreted with caution. They provide an initial characterization of the stochastic properties of the series without allowing definitive conclusions to be drawn regarding the presence of long memory. To resolve this ambiguity, the analysis is complemented in the following section by estimating the fractional differencing parameter using several semi-parametric estimators based on the low-frequency behavior of the spectrum.
\begin{table}[htbp]
	\footnotesize
	\caption{\label{tab:stationarity_tests}
		Results of the ADF and KPSS Stationarity Tests for the Logarithmic Monthly White Maize Prices.}
	\centering
	
	\begin{tabular}{lcccc}
		\hline
		\textbf{Market} &
		\textbf{ADF Statistic} &
		\textbf{ADF $p$-Value} &
		\textbf{KPSS Statistic} &
		\textbf{KPSS $p$-Value} \\
		\hline
		
		Korbongou & -4.030 & 0.0013 & 1.116 & $<0.01$ \\
		Cinkassé  & -2.841 & 0.0527 & 0.751 & $<0.01$ \\
		Kara      & -3.269 & 0.0163 & 1.427 & $<0.01$ \\
		Anié      & -1.880 & 0.3416 & 1.254 & $<0.01$ \\
		Amégran   & -3.033 & 0.0319 & 0.708 & 0.0128 \\
		Lomé      & -2.173 & 0.2163 & 0.888 & $<0.01$ \\
		
		\hline
	\end{tabular}
	
	\vspace{0.15cm}
	
	\begin{minipage}{0.92\textwidth}
		\footnotesize
		\textit{Note:} The null hypothesis of the ADF test is the presence of a unit root,
		whereas the null hypothesis of the KPSS test is stationarity around a constant.
		The tests are performed on the logarithmic monthly price series.
	\end{minipage}
	
\end{table}
\subsection{Estimation of the long-memory parameter}

The long memory of the logarithmic monthly white maize price series was estimated using the four semi-parametric estimators presented in the methodology section, namely the GPH estimator, the Local Whittle estimator, the Exact Local Whittle (ELW) estimator, and the wavelet log-regression estimator. Although these approaches rely on different estimation principles, they all exploit the asymptotic behavior of the spectrum at low frequencies. Their comparison makes it possible to assess the robustness of the estimated values before selecting the most appropriate estimator for fractional modeling.

Table~\ref{tab:memory_estimators} reports the estimates of the memory parameter $d$ obtained for the six markets under study.
\begin{table}[htbp]
	\footnotesize
	\caption{\label{tab:memory_estimators}
		Comparison of Semi-Parametric Estimates of the Long-Memory Parameter $d$ for the Logarithmic Monthly White Maize Prices.}
	\centering
	
	\begin{tabular}{lcccc}
		\hline
		\textbf{Market} &
		\textbf{GPH} &
		\textbf{Local Whittle} &
		\textbf{Exact Local Whittle} &
		\textbf{Wavelet Log-Regression} \\
		\hline
		
		Korbongou & 0.670 & 0.587 & 0.602 & 0.609 \\
		Cinkassé  & 0.756 & 0.670 & 0.700 & 0.817 \\
		Kara      & 0.729 & 0.637 & 0.559 & 0.635 \\
		Anié      & 0.681 & 0.489 & 0.425 & 0.571 \\
		Amégran   & 0.555 & 0.476 & 0.434 & 0.515 \\
		Lomé      & 0.671 & 0.602 & 0.504 & 0.558 \\
		
		\hline
	\end{tabular}
	
	\vspace{0.15cm}
	
	\begin{minipage}{0.92\textwidth}
		\footnotesize
		\textit{Note:} Estimates of the fractional differencing parameter obtained using
		the Geweke--Porter--Hudak (GPH), Local Whittle, Exact Local Whittle, and
		wavelet log-regression estimators for the logarithmic monthly white maize price series.
	\end{minipage}
	
\end{table}

Overall, the four methods provide consistent estimates of the memory parameter, although differences in magnitude are observed across estimators and markets. All the estimates are strictly positive, providing a first indication of the existence of long-range dependence in the logarithmic white maize price series. This consistency across methods based on different theoretical foundations strengthens the credibility of the long-memory diagnosis.

The GPH and Wavelet log-regression estimators generally produce the highest estimates, whereas the Local Whittle and Exact Local Whittle estimators yield more conservative estimates. This behavior is consistent with the well-known properties of these methods: the GPH estimator is more sensitive to the choice of the frequency band, whereas Whittle-type estimators generally exhibit greater statistical efficiency and lower finite-sample variance.

Among the four estimators considered, the analyses presented in the remainder of this study are based on the Exact Local Whittle estimator. This choice is motivated by its favorable asymptotic properties, its consistency for both stationary and non-stationary processes, and its good performance widely documented in the literature \cite{shimotsu2005}.

The estimates obtained using the Exact Local Whittle estimator range from 0.425 for the Anié market to 0.700 for the Cinkassé market. The Korbongou (0.602), Cinkassé (0.700), Kara (0.559), and Lomé (0.504) markets exhibit values greater than 0.5, suggesting strong temporal persistence consistent with a non-stationary fractionally integrated process. In contrast, the Anié (0.425) and Amégran (0.434) markets exhibit estimates between 0 and 0.5, indicating the presence of long memory while remaining stationary in the covariance sense.

These results highlight a marked persistence of shocks affecting the logarithmic white maize price series across all the markets considered. In particular, disturbances observed in the markets exhibiting the highest values of $d$ tend to dissipate slowly over time, reflecting strong inertia in price dynamics. This property justifies the use of fractionally differenced models to adequately represent the long-range dependence of the logarithmic series. The performance of these models is examined in the following section.
\subsection{Model selection and comparison}

The objective of this analysis is to determine whether accounting for long memory improves the modeling of the logarithmic monthly white maize price series compared with conventional short-memory approaches. Three families of models were therefore fitted independently to each of the six logarithmic price series: SARIMA models, representing the conventional approach with a seasonal component; ARFIMA models incorporating fractional differencing; and SARFIMA models combining long memory and seasonality.

To ensure a homogeneous comparison among the three model families, the autoregressive and moving average orders were determined automatically using a systematic model selection procedure. For each family, several admissible specifications were estimated and subsequently compared using the Bayesian Information Criterion (BIC). The fractional differencing parameter $d$ of the ARFIMA and SARFIMA models was fixed at the value obtained from the Exact Local Whittle estimator presented in the previous section. The specification retained for each family therefore corresponds to the model with the lowest BIC value, a criterion preferred because of its stronger penalty on model complexity and its ability to favor the most parsimonious specifications.

Table~\ref{tab:model_comparison} summarizes, for each market, the best specification obtained within each model family together with the corresponding BIC value. The minimum BIC values are shown in bold to highlight the model selected for each market.

\begin{table}[htbp]
	\footnotesize
	\caption{\label{tab:model_comparison}
		Comparison of the Competing Time Series Models Fitted to the Logarithmic Monthly White Maize Prices.}
	\centering
	
	\begin{tabular}{llcc}
		\hline
		\textbf{Market} & \textbf{Model} & \textbf{Specification} & \textbf{BIC} \\
		\hline
		
		Korbongou & SARIMA  & $(0,1,1)(1,0,1)_{12}$ & \textbf{-222.910} \\
		& ARFIMA  & $(1,d,0)$               & -221.966 \\
		& SARFIMA & $(1,d,0)(0,0,0)_{12}$   & -221.966 \\
		\hline
		
		Cinkassé  & SARIMA  & $(0,1,0)(1,0,1)_{12}$ & \textbf{-364.348} \\
		& ARFIMA  & $(1,d,0)$               & -354.136 \\
		& SARFIMA & $(1,d,0)(0,0,0)_{12}$   & -354.136 \\
		\hline
		
		Kara      & SARIMA  & $(0,1,0)(0,0,0)_{12}$ & -289.897 \\
		& ARFIMA  & $(1,d,0)$               & \textbf{-293.376} \\
		& SARFIMA & $(1,d,0)(0,0,0)_{12}$   & \textbf{-293.376} \\
		\hline
		
		Anié      & SARIMA  & $(1,1,2)(1,0,1)_{12}$ & -180.967 \\
		& ARFIMA  & $(1,d,0)$               & -155.606 \\
		& SARFIMA & $(1,d,0)(1,0,1)_{12}$   & \textbf{-190.131} \\
		\hline
		
		Amégran   & SARIMA  & $(0,1,1)(0,0,0)_{12}$ & -102.334 \\
		& ARFIMA  & $(2,d,0)$               & \textbf{-107.254} \\
		& SARFIMA & $(2,d,0)(0,0,0)_{12}$   & \textbf{-107.254} \\
		\hline
		
		Lomé      & SARIMA  & $(1,1,1)(1,0,1)_{12}$ & -298.829 \\
		& ARFIMA  & $(1,d,0)$               & -293.469 \\
		& SARFIMA & $(1,d,0)(1,0,0)_{12}$   & \textbf{-301.958} \\
		\hline
	\end{tabular}
	
	\vspace{0.15cm}
	
	\begin{minipage}{0.92\textwidth}
		\footnotesize
		\textit{Note:} The preferred specification corresponds to the smallest
		Bayesian Information Criterion (BIC). The selected BIC values are highlighted in bold.
	\end{minipage}
	
\end{table}
The results reported in Table~\ref{tab:model_comparison} reveal substantial heterogeneity in the dependence structures across the markets considered. No single model family systematically outperforms the others, suggesting that the temporal dependence mechanisms of the logarithmic white maize price series differ markedly from one market to another.

The Korbongou and Cinkassé markets are the only ones for which the SARIMA model yields the lowest BIC. Despite the relatively high estimates of the long-memory parameter obtained previously, the introduction of fractional differencing does not sufficiently improve the quality of the fit to offset the penalty associated with the additional model complexity. These two series therefore appear to be adequately represented by a short-memory dynamics combined with an annual seasonal component.

In contrast, the Kara and Amégran markets are better described by fractionally integrated models. The ARFIMA and SARFIMA models yield exactly the same BIC value for these markets. This equality is explained by the fact that the seasonal component estimated in the SARFIMA model is equal to zero, making it empirically equivalent to the corresponding ARFIMA model. These results indicate the presence of long-range dependence without any additional detectable seasonal effect.

The Anié market exhibits a different pattern. The SARFIMA model achieves a substantial improvement in BIC compared with the SARIMA and ARFIMA models, indicating that the combination of fractional differencing and a seasonal component considerably improves the representation of the series. A similar result is observed for the Lomé market, where the SARFIMA model yields the lowest BIC value among the three model families considered.

These results naturally extend the exploratory analyses presented previously. The autocorrelation functions revealed a slow decay of the correlations, a characteristic of strong temporal persistence, while the periodograms showed a high concentration of spectral power at low frequencies. The semi-parametric estimators also yielded long-memory parameter estimates ranging from 0.425 to 0.817 depending on the market and the estimation method. The model comparison confirms that this persistence has practical implications for modeling: for four of the six markets considered, the introduction of fractional differencing effectively improves the model fit according to the BIC criterion.

Overall, these results show that the long-memory hypothesis is empirically relevant for the majority of the markets considered, while revealing heterogeneity in the dependence structures across markets. They therefore motivate the consideration of fractionally differenced models, whose performance is compared with that of conventional short-memory models. The following section is devoted to the residual diagnostic analysis of the selected models in order to verify that the temporal dependence structure of the logarithmic price series has been adequately represented.
\subsection{Diagnostic analysis of the selected models}

After selecting the best model for each market using the Bayesian Information Criterion (BIC), it is essential to verify that the temporal dependence present in the series has been adequately captured. To this end, we analyze the residuals of the selected models using the Ljung--Box test proposed by \cite{ljung1978}, together with their empirical autocorrelation functions.

Table~\ref{tab:residual_diagnostics} reports the \emph{p}-values of the Ljung--Box test computed at lags 12, 24, and 36. Under the null hypothesis, the residuals exhibit no significant autocorrelation. The values shown in bold correspond to cases where this hypothesis is rejected at the 5\% significance level.

For the Korbongou, Cinkassé, and Anié markets, the three \emph{p}-values remain above the significance level, indicating that the selected models produce residuals consistent with white noise. These results suggest that both the short-term dependence and the long-memory structure have been adequately captured by the selected models.

In contrast, for Kara, Amégran, and Lomé, at least one of the Ljung--Box statistics leads to the rejection of the null hypothesis of no residual autocorrelation. This observation indicates that a weak dependence structure remains in the residuals despite fitting fractionally integrated models. Nevertheless, these models remain those providing the best compromise between goodness-of-fit and parsimony according to the BIC criterion.

\begin{figure}[htbp]
	\centering
	\makebox{\includegraphics[width=\textwidth]{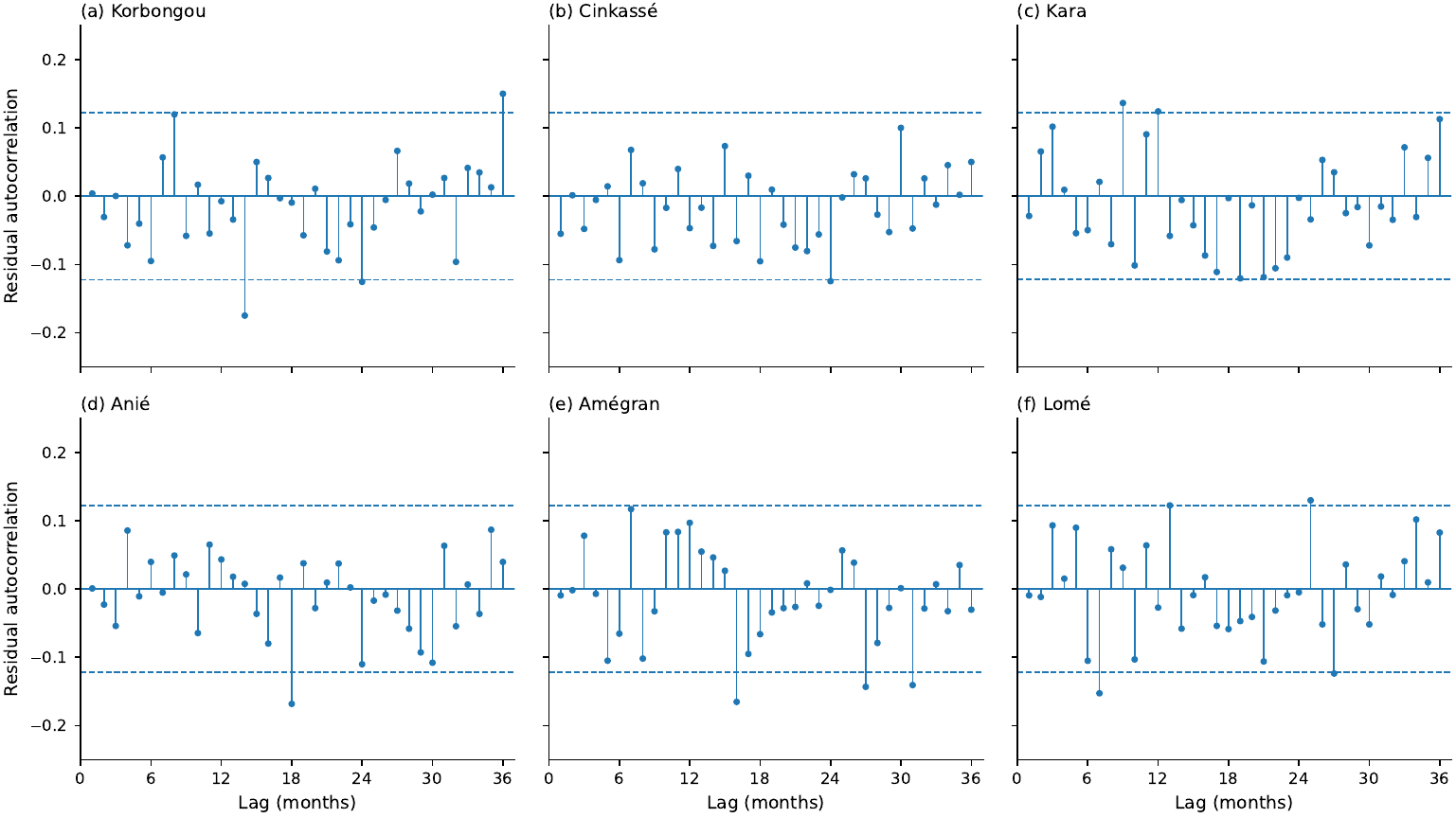}}
	\caption{\label{fig:residual_acf_selected_models}
		Sample Autocorrelation Functions of the Residuals from the Models Selected by the Bayesian Information Criterion for the Six Markets. The Dashed Horizontal Lines Represent the Approximate 95\% Confidence Bounds under the Hypothesis of Zero Residual Autocorrelation.}
\end{figure}

Figure~\ref{fig:residual_acf_selected_models} complements this analysis by displaying the autocorrelation functions of the residuals from the selected models. For Korbongou, Cinkassé, and Anié, the residual autocorrelations generally fluctuate around zero and remain mostly within the 95\% confidence bands, thereby confirming the conclusions of the Ljung--Box test. In contrast, the Kara, Amégran, and Lomé markets still exhibit a few spikes that occasionally exceed the confidence limits, indicating the persistence of slight residual autocorrelation. However, these results remain localized and do not call into question the overall adequacy of the selected models.

Overall, the residual diagnostics show that the selected models provide a satisfactory representation of the dependence structure of the logarithmic white maize price series for the majority of the markets considered. The few residual dependencies observed in some markets nevertheless suggest that these series exhibit a complex temporal dynamics, potentially influenced by exogenous factors or structural changes that are not fully accounted for by the models considered.

\begin{table}[htbp]
	\footnotesize
	\caption{\label{tab:residual_diagnostics}
		Residual Diagnostics of the Selected Models Based on the Ljung--Box Test.}
	\centering
	
	\begin{tabular}{lcccc}
		\hline
		\textbf{Market} &
		\textbf{Selected Model} &
		\textbf{LB(12)} &
		\textbf{LB(24)} &
		\textbf{LB(36)} \\
		\hline
		
		Korbongou & SARIMA &
		0.2801 &
		0.0767 &
		0.1023 \\
		
		Cinkassé & SARIMA &
		0.6455 &
		0.3430 &
		0.6243 \\
		
		Kara & ARFIMA &
		\textbf{0.0469} &
		\textbf{0.0137} &
		\textbf{0.0403} \\
		
		Anié & SARFIMA &
		0.6506 &
		0.4219 &
		0.4113 \\
		
		Amégran & SARFIMA &
		\textbf{0.0431} &
		0.0672 &
		\textbf{0.0457} \\
		
		Lomé & SARFIMA &
		\textbf{0.0391} &
		0.1049 &
		0.0571 \\
		
		\hline
	\end{tabular}
	
	\vspace{0.15cm}
	
	\begin{minipage}{0.92\textwidth}
		\footnotesize
		\textit{Note:} The table reports the $p$-values of the Ljung--Box test
		computed on the residuals of the selected models. Values in bold indicate
		rejection of the null hypothesis of no residual autocorrelation at the
		5\% significance level.
	\end{minipage}
	
\end{table}
Overall, the analyses conducted show that the logarithmic white maize price series exhibit significant temporal persistence, the intensity of which varies across markets. They also show that fractionally integrated models do not systematically provide the best performance, thereby highlighting the importance of a rigorous data-driven model selection procedure.

\section{Conclusion}
\label{sec:conclusion}

This study characterizes the long memory of logarithmic monthly white maize price series observed in the major markets of Togo and evaluates the relevance of fractionally integrated models for their statistical modeling. A progressive framework was implemented, combining exploratory analysis of the series, stationarity testing, long-memory estimation using several semi-parametric estimators, and a comparison of the performance of SARIMA, ARFIMA, and SARFIMA models.

The results show that the logarithmic price series exhibit characteristics consistent with long-range dependence. This property is consistently revealed by the autocorrelation functions, the periodograms, and the four long-memory estimation methods considered in this study. Among these methods, the Exact Local Whittle estimator provides stable estimates consistent with the theoretical properties of fractionally integrated processes and was therefore retained for the subsequent modeling stage.

The comparison of SARIMA, ARFIMA, and SARFIMA models shows that accounting for long memory improves the statistical representation of the series for several markets, although it does not constitute a universal solution. While fractionally integrated models are selected for the majority of the markets considered, SARIMA models remain the preferred specifications for some markets. The main finding of this study is therefore that the identification of long memory is not, by itself, sufficient to justify the systematic use of a fractionally integrated model. Rather, the presence of long memory and the selection of the most appropriate time series model should be regarded as two related but distinct statistical questions. Model selection should rely on a rigorous empirical comparison among competing model families while accounting for the specific characteristics of each series.

The residual diagnostics indicate that the selected models generally capture the dependence structure of the logarithmic price series. Nevertheless, the residual dependencies observed for some markets suggest that agricultural price dynamics may also reflect economic, climatic, institutional, or other factors that are not explicitly accounted for by the univariate models considered.

Beyond the empirical results, this study makes several contributions. To the best of our knowledge, it is the first study devoted to the characterization of long memory in white maize prices across the major markets of Togo. It also provides a systematic comparison of four semi-parametric estimators of the long-memory parameter (GPH, Local Whittle, Exact Local Whittle, and wavelet log-regression), followed by a comparative evaluation of SARIMA, ARFIMA, and SARFIMA models within a unified methodological framework. More importantly, the results empirically demonstrate that evidence of long memory does not necessarily imply the superiority of fractionally integrated models. Long-memory identification therefore provides useful information about the dependence structure of agricultural price series, but should be complemented by model selection based on objective criteria and residual diagnostics. The proposed methodological framework is fully reproducible and can be applied to other agricultural markets or commodities exhibiting persistent temporal dependence.

\end{document}